\documentclass[useAMS,usenatbib]{mn2e}
\usepackage{epsfig}
\usepackage{natbib}
\usepackage{color}
\usepackage{soul}
  \usepackage{aas_macros}

\newcommand{\msun}{$M_{\odot}$}

\newcommand{\bevii}{$^{7}$Be}

\newcommand{\lii}{Li\,{\sc i}}

\newcommand{\livii}{$^{7}$Li}
\newcommand{\liviii}{$^{7}$Li\,{\sc i}}

\newcommand{\caii}{Ca\,{\sc ii}}

\newcommand{\iiihe}{$^{3}$He}

\title[ \livii\  evolution  ]{   \livii\  evolution in the  thin and thick disks of the Milky Way}
\author[]{ G. Cescutti$^{1}$  
 \thanks{E-mail: gabriele.cescutti@inaf.it (GC)} 
 and P. Molaro $^{1}$, \\
$^{1}$  INAF-Osservatorio Astronomico di Trieste, Via G.B. Tiepolo 11, I-34143 Trieste, Italy\\
}
\begin{document}

\date{Accepted.... Received 2016-07-31}

\pagerange{\pageref{firstpage}--\pageref{lastpage}} \pubyear{2002}

\maketitle

\label{firstpage}

\begin{abstract}
 
  Recent detection of the isotope \bevii\ (that decays into \livii )
  in the outburst of classical novae confirms the suggestion made in
  the 70s that novae could make \livii.  We reconsidered the role of
  novae as producers of \livii~ by means of a detailed model of the
  chemical evolution of the Milky Way.  We showed that novae could be
  {\it the} Galactic \livii~ source accounting for the observed
  increase of Li abundances in the thin disk with stellar
    metallicity. The best agreement with the upper envelope of the
    observed Li abundances is obtained for a delay time of $\approx$ 1
    Gyr for the nova production and for an effective \livii~ yield of
    $1.8 (\pm 0.6) \times 10^{-5}$ M$_{\sun}$ over the whole nova
    life. Lithium in halo stars is depleted by $\approx$ 0.35 dex,
    assuming the pristine abundance from standard big bang
    nucleosynthesis. We elaborate a model that matches the pristine
    stellar abundances, assuming that all stars are depleted by 0.35 dex.
    In this case, the delay remains the same, but the Li yields are
    doubled compared to the previous case. This model also has the
    merit to match the Li abundance in meteorites and young TTauri
    stars.  The thick disk model, adopting the parameters derived for
    the thin disk, can also explain the absence of increase of Li
    abundances in its stars.  The thick disk is old, but formed and
    evolved in a time shorter than that required by novae to
    contribute significantly to \livii.  Therefore, no \livii ~
    enhancement is expected in thick disk stars.  We show that the
    almost constant Li abundance in the thick disk results from
    the compensation of stellar astration by spallation processes.
   
\end{abstract}

\begin{keywords}
{stars: individual V5668 Sgr; stars: novae
-- nucleosynthesis, abundances; Galaxy: evolution -- abundances}
\end{keywords}

\section{Introduction}

\livii\ nuclei are the heaviest ones produced in significant
amounts during the Big Bang. The primordial \livii\ production is a
sensitive function of the baryon-to-photon ratio and can be estimated
in the framework of standard primordial nucleosynthesis once the
baryon density is taken either from the primordial deuterium abundance
or from the fluctuations of the cosmic microwave background (CMB). The
expected primordial value is of A(Li) = 2.6 which is a factor of three
to four higher than what measured in the halo dwarf stars
\citep{Spite82}.  This discrepancy is normally referred to as the {\it
  Cosmological Lithium problem}. The problem became clear by the
  time WMAP measurement of universal baryon density was used to infer
  the primordial Li abundance within the standard primordial
  nucleosynthesis \citep{Cyburt03}, although there were already earlier hints
  \citep{Cayrel98}.  More recently, the baryon density
  measured either by the Planck mission \citep{Planck} or by 
  deuterium in quasar spectra confirmed the disagreement with the
  lithium abundance measured in the halo stars \citep{Fields14}.  The
origin of this discrepancy has still to be identified with confidence,
but a possible stellar fix has been recently proposed by \citet{Fu15}.
Since the present Li abundance as measured in meteorites
\citep{Krank64} or in the young TTauri stars is A(Li) $\approx$ 3.3, a
Galactic source is required to account for the increase from the
initial value of 2.6.  The identification of these sources is
generally referred to as the {\it Galactic Lithium problem}.  A large
variety of nucleosynthesis processes and sources was suggested so far.  An established source is the
spallation of atoms in the interstellar-medium by energetic cosmic
rays. The same processes make also $^6$Li, $^9$Be and $^{10,11}$B.
Thus, from the abundance of these isotopes, in particular $^9$Be, and
by means of the relative cross-sections it is possible to estimate the
fraction of \livii\ produced by spallation. Spallation processes
integrated along the Galactic life can account for at most  30\% of the
presently measured \livii\ and require other source(s) to enrich the
Galaxy to the present values. Other proposed sources are the
spallation processes in the flares of low mass active stars, red
giants (RG), asymptotic giant branch (AGB) stars, novae and neutrino
induced nucleosynthesis in supernovae explosions.  \livii ~ has been
observed to be greatly enhanced in some AGB stars
 \citep{SmithLambert89,SmithLambert90} and the
case of Li in red giants was nicely reviewed in \citet{Casey16}.
  
Different sources have different time scales for the production and
therefore the rate of lithium increase is a different way to asses
their relative contributions. However, the possible \livii\
destruction inside a star complicates the picture. For instance, the
Sun shows A(Li) = 1.04 which is more than two orders of magnitudes
lower than the proto-solar nebula due to internal and poorly
understood destruction.  During the stellar evolution off the
  main sequence surface layers with lithium are mixed with more
  internal ones producing a Li diluition which can be nicely observed
  in the globular clusters, such as NGC 6397 \citep[e.g.][]{Lind09}. 

Observations of metal poor stars show that \livii ~ remains constant
at A(Li) $\approx$ 2.3 from the lowest metallicities up to
approximately [Fe/H] $\approx -$1 when the abundance starts to rise
reaching the meteoritic value of A(Li) =3.3. at solar
metallicities. After \citet{reb88} it became common to assume the
upper envelope of the distribution of  \livii \, abundances to
trace the \livii\ abundance evolution.  Galactic chemical evolution
(GCE) models of \livii\ were pioneered by \citet{D'Antona91} and
\citet{rom99} where the different contributions and time scales were
discussed.  In these models a shared feature is that the AGB or SNe do
not contribute much to the \livii\ production, and novae are 
potentially an important source.  \citet{Matteucci95} also studied the
possible contribution of $\nu$ -process during SNe II explosions.
 The interplay of different sources was studied in the context of
 the GCE model of lithium by \citet{Travaglio01}, whereas
  \citet{Alibes02} analysed in particular the role of spallation for
  lithium, beryllium and boron.  \citet{Prantzos12} concluded that
  30\% of the lithium content of our Galaxy at most can be produced by
  galactic cosmic rays, and a stellar origin is needed for the
  remaining fraction. \citet{Romano01} and \citet{Prantzos17}
  identified low-mass giant stars as the best candidates for
  reproducing the late rise off the lithium-metallicity plateau. 

In fact, about 1-2 \% of red giants show Li higher than A(Li)
  $\approx$ 1.5 that is the expected value predicted by standard
  stellar evolution theory
due to evolution along the first giant branch
  \citep{Casey16,Smiljanic18,Lyubimkov16}.  The most Li-rich giant
  recently discovered by \citet{Yan18} reaches the value of A(Li) =
  4.51 $\pm 0.09$. However, only about 30 other red giants show A(Li)
  greater than the meteoritic value, i.e. only about 0.1\% of the red
  giants studied. The most promising explanation for this Li
  enrichment is similar to the Cameron - Fowler mechanism
  \citep{cam71} proposed to explain the Li in AGB stars. However, in
  RG stars the introduction of the  extra-mixing mechanism is required to
  bring fresh lithium to the surface
  \citep{Sackmann99,Lattanzio15}. Another possible explanation is that
  the lithium comes from external pollution, for example in the case
  of an engulfment of planets or planetesimals during the evolution to
  the red giant branch \citet{Siess99}. Whatever the origin the high
  A(Li) cannot be maintained for long time due to convection activity
  in these stars as also shown by the low percentage of
  RG which are Li rich \citep{Yan18}. Therefore, their contribution to the Li
  enrichment of the interstellar medium remains quite uncertain.

  The significant increase of the measurements provided by recente
  surveys revealed a different behaviour in the \livii\ evolution
  between the thin and thick disk of the Milky Way. At the lower
  metallicity end the thick and thin disk stars show the same Li
  abundance \citep{Molaro97}. At higher metallicities the increase in
  Li was steeper in the thin than the thick disks by
  \citet{Guiglion16} and \citet{Fu18}. The precise behaviour of
  \livii\ abundance in the thick disk  as a function of 
    metallicity is still matter of debate. It was found constant and
  approximately at the same value of the most metal poor stars by
  \citet{Ramirez12}, decreasing by \citet{Delgado15} and
  \citet{Bensby18} and slightly increasing by \citet{Fu18}.  The
  separation between thin and thick stars is not always
  straightforward and the presence of some contamination could explain
  the slightly different results for the thick disk.  In any case, all
  these analyses show that the thick disk abundances are equal or
  lower than the thin disk ones of similar metallicities.  An
  additional feature for thin disk stars emerged recently: Li
  abundances of stars above solar metallicity show lower values than
  at solar metallicity suggesting a puzzling decrease of
  lithium. \citet{Guiglion16} and  \citet{Prantzos17} tried to explain
  these features in terms of stellar migration as it is has been done
  for the similar [O/Fe] versus [Fe/H] behaviour, but without
  achieving a robust conclusion.

Thermo-nuclear production of \bevii\ (that decays into \livii ) during
nova explosions were proposed by \citet{arn75} and \citet{sta78}.
At temperatures of ~ 150 million K \bevii\ is formed from the reaction
\iiihe\ ($\alpha$, $\gamma$)\bevii ~ \citep{her96}.  To avoid
destruction \bevii\ needs to be carried to cooler regions by
convection on a short time scale as in the Cameron-Fowler mechanism
\citep{cam71}.  When these cooler regions are subsequently ejected,
\bevii\ could be observed in absorption in the nova outburst
\citep{jos98}.  Predicted in the mid 70s, observational evidence of Li
in a nova outburst has been found only recently by \citet{izz15} who
reported the possible detection of the \liviii ~ $\lambda\lambda$6708
line in the spectra of Nova Centauri 2013 (V1369 Cen). This has been
followed by the detection of the mother nuclei $^{7}$Be in the
post-outburst spectra of classical novae by \citet{taj15},
\citet{taj16}, \citet{Molaro16}, \citet{ Izzo18} and
\citet{Selvelli18}.  The observed \bevii\ decays into \livii\ with a
mean-life of 53 days and all \bevii\ observed ends up into \livii.
The relative abundance of \bevii/H has been measured in four classical
novae with overproduction factors which range between 4 and 5 orders
of magnitude over the meteoritic abundance.
 In this paper, we revise the \livii\ evolution by
considering this new evidence and see if we can reproduce the \livii~
behaviour. We also explore whether a Li synthesis by classical novae
could also explain the different \livii\ behaviours observed in the
thick disk of the Galaxy.


\section{\livii\ Galactic Chemical evolution of the thin disk}

\subsection{Observational Data for the thin disk}\label{data_thin}

The abundances for the Galactic thin disk stars are taken from the
Ambre project \citep{Guiglion16} and from the data by
\citet{Bensby18}.  In the Ambre project \citep{Guiglion16} lithium
abundances for 7272 stars were derived. However, we use only their
working sample of 3009 stars that comprises only dwarf stars.
\citet{Bensby18} measured Li abundances for 420 dwarf stars and
provided upper limits on the Li abundance for a further 121.  We
integrate the results of this data set with previous results
\citep{Bensby14} for stellar ages, temperature and stellar kinematics,
as defined by the calculated probability functions.  As shown in
\citet{Bensby18}, some stars are not correctly classified as
  thick or thin disk by a chemical selection only, although it is
successful in most cases. In fact, depending on alpha element the
quality of the chemical selection changes significantly. Tiny
differences in the chemical selection can change the outcome quite
dramatically and the selection becomes increasingly difficult toward
solar metallicity.  We therefore consider the kinematical selection as
the best one, and we use the chemical selection only when the
kinematical information is not present.  The kinematical parameters
are not available in the Ambre data and therefore we use their
chemical selection with 2671 thin disk stars.
For the Ambre data we compute an upper envelope obtained by computing
the mean over the five data points with the highest \livii\ over 10
bins of 0.1 dex in [Fe/H] and after clipping for the outliers. The
envelope is shown in Fig \ref{fig1} with the shadowed area covering
one standard deviation from the mean value.  For \citet{Bensby18} data
we do not compute the envelope due to the small number of stars and
show only stars with effective temperature higher than 5900 K to
partially exclude the main-sequence stars where internal stellar
depletion is occurring.  This temperature threshold is derived
  from Fig. 8 in \citet{Bensby18} and Fig. 11 in \citet{Lind09}. With
  this temperature cut off the sample is reduced to 116.
To note that in general for stars with lower Li abundances it is
unclear whether \livii\ was partially depleted by convections or
similar mechanisms or reflects a truly initial lower abundance.

\subsection{Observational constraints to novae \livii\  yields}

Normally \liviii~ is not detected in the spectra of novae outbursts
and the first, and so far unique, \lii\ detection by \citet{izz15}
implies that the physical conditions in the ejecta of post-outburst
novae only rarely permit the survival of neutral \liviii. However,
the recent detection of mother nuclei \bevii\ in the post-outburst
spectra of all the novae where it was possible to study the
presence of \bevii~ so far shows that thermonuclear production of
\bevii\ is effectively taking place and is probably a common feature of
classical novae.

 For three novae the abundance of \bevii\ is estimated relatively to
 the \caii\, for unsaturated and resolved lines which are assumed to
 represent the abundances in the whole material ejected.  The
 \bevii/H by number derived from the observations and corrected for
 the \bevii\ decay assuming all the \bevii\ has been made around the
 nova maximum.  In Nova Herculis 1990 the abundance was also derived
 from the emission which takes into account the whole envelope
 and found of $\approx$ 2 $\cdot$ 10$^{-5}$, consistent with the 
abundance derived from absorption lines
 \citep{Selvelli18}.  To note that the observed yields are
 significantly larger than the maximum theoretical yields predicted by
 the models of \citet{ jos98}.  On the long term \bevii\ $=$\livii\
 and the derived fractions correspond to a \livii ~ logarithmic
 overabundances of 4 to 5 dex with respect to the meteoritic value of
 1.3 $\cdot$ 10$^{-9}$ \citep{lod09}.

These overproduction factors imply a production factor of
   $ 1-10 \cdot$ 10$^{-9}$ \msun \, per nova event assuming an ejected
   mass of $\approx$ 10$^{-5}$ \msun~ and a total production of
   $ 1-10 \cdot$ 10$^{-9}$ \msun~ of Li during the whole nova life
   assuming a typical number of bursts of 10$^4$. 
  A rate of 20 yr$^{-1}$ of nova
events in a Galaxy lifetime of $\approx$ 10$^{10} $ yr producing
$M_{Li}$ $\approx$  3 $\cdot$ 10$^{-9}$ \msun~ is enough to produce
$M_{Li}$ $\approx$ 600 \msun which is comparable to that
estimated to be present in the Milky Way.  This simple estimation
shows that novae could indeed be one of the main factories of \livii ~
in the Galaxy.

Therefore, we decide to reconsider the role of novae as main producers
of \livii~ in the Milky Way.  To investigate this role, we implement
the lithium production from novae in a detailed GCE model of the Milky
Way thin disk (see next Section).  We do not assume the
observational yields into the modelling  directly, given the spread in the
observed abundances. We rather estimate the best set of parameters for
lithium production by novae - described in the next Section -comparing
the chemical evolution model results to the abundance measured in thin
disk stars.  Only at the end of this procedure we verify whether or
not these theoretically computed yields are compatible with the
observational constraints obtained from novae outbursts.

\subsection{  Thin Disk chemical evolution model}
 
Our model for the thin disk adopts prescriptions similar to that used
in \citet{Romano10} and \citet{Grisoni17}.  \citet{Grisoni17} modified
the original two infall frameworks \citep{Chiappini97} and set the
parameters to reproduce the $\alpha$ elements and the metallicity
distribution function determined in the AMBRE project at best.  The
initial mass function (IMF) is from \citet{Kroupa01}, whereas the
stellar lifetime are from \citet{MM02}.  SNe Ia follows the single
degenerate scheme as \citet{Matteucci86} with a fraction of successful
binary systems of 0.05 and iron yields from W7 model by
\citet{Iwamoto99}.  Iron nucleosynthesis for massive stars is from
\citet{Kobayashi11}.  The main characteristic of the model is a
prolonged exponential infall following this relation:

\begin{equation}
G_{inf.}(t)=A\cdot e^{(t/\tau_{D})}.
\end{equation}

with timescale of $\tau_{D}$ = 4 Gyr  and a total  evolution which lasts for 10 Gyr. 

The star
formation follows this equation:
\begin{equation}
\psi(r,t)=\nu\left(\frac{\Sigma(r,t)}{\Sigma(r_{\odot},t)}\right)^{2(k-1)}
\left(\frac{\Sigma(r,t_{Gal})}{\Sigma(r,t)}\right)^{k-1}G^{k}_{gas}(r,t).
\end{equation}
with a mild efficiency ($\nu_{SFR}$=0.5). 
\subsubsection{ Lithium nucleosynthesis}
In this paper we assume that the principal producers of \livii\ are
nova systems. Given the relatively large spread in the lithium
produced by novae, we study the space of the parameters that describes
this production and we compare the results obtained by means of a
detailed chemical evolution model to observational data of thin disk
stars.  The main assumptions are the following:

\begin{itemize}

\item Only binary systems formed by stars in the mass range
  0.8\msun$ <$ M $<$ 8\msun \, can develop nova systems that produce
  \livii \, and the main parameter is the fraction $N_{lithium}$ of
  these systems that actually develop nova systems and produce \livii.
  The probability to form a binary system of a certain mass is
  weighted on the IMF as it is made of a single star with the same
  mass.  Given our not perfect knowledge of the coupling between
    IMF and binaries, this method is a reasonable approximation and
  is similar to that developed in \citet{Matteucci86} for the progenitors
  of SNe Ia.  The maximum total mass of the binary is therefore
  16\msun.  On the other hand, the minimum mass ($M_{low}$) of the
  binary would be 1.6\msun; however, once we explored this parameter we
  obtained better results using $M_{low}$=3\msun. A model with
  $M_{low}$=1.6\msun~ is presented to show the different outcome.  The
  probability of configurations of primary and secondary stars are
  considered following this equation:
\begin{equation}
f(\mu)=2^{1+\gamma}(1+\gamma)\mu^{\gamma}  
\end{equation}
where $\mu$ is the mass of the secondary divided by the total mass of
the binary system and $\gamma=2$ \citep{Greggio83}.  In this way,
  contrary to most of the previous chemical evolution models
  considering lithium production from novae, we find a 
  probability distribution of the mass of the secondary star.

\item $^7$Li production  takes place after the time required  by  the primary
  to evolve into a white dwarf plus a delay  time $\tau_{nova}$  which is the time
  the white dwarfs need to accumulate material to ignite the first nova
  outburst.  Binaries with  components of the same mass 
    never develop a nova  and, therefore, we consider  \livii\ producers only binaries
  with sufficient difference between the lifetimes of primary and
  secondary, namely $\tau(M_{sec})-\tau(M_{prim}) \ge \tau_{nova}$.

\item Nova systems encounter several bursts along their whole life and
  a typical number is 10$^{4}$ bursts \citep{Bath78,Shara86}.  In
  previous chemical evolution models, for mathematical and technical
  simplification a single lithium production event was considered to
  make the whole lithium of an entire life.   In our work we
    improve this treatment and we consider multiple ejections instead
    of ejecting all the produced \livii\ in as single event, as
    previously done.  We find significant differences after
    considering more than a single burst.  Ideally, we should take
    into account 10$^{4}$ bursts, but assuming an increasing number of
    ejections (up to 100), we note that these differences become
    negligible with more that five enrichments and to spare
    computational time we use five ejection events for each nova.

  \item For the sake of simplicity, we assume that during their life
    all novae produce the same amount of \livii\ and in all events
    regardless of the masses of the original binary systems, or the
    $\tau_{nova}$ considered. We define this quantity $^{Li}Y_{Nova}$
    and, as we will show in the next Section, the best value we
    find is $^{Li}M_{Nova}=1.8 \times 10^{-5}$ \msun.  This simplified
    assumption is contradicted by present observations which show one
    order of magnitude difference in the novae where \bevii\,
    i.e. \livii has been measured. On the other hand, we prefer to
    keep the model as simple as possible. Moreover, this value is
    compatible with the observations and can be taken as a typical
    averaged amount.

\end{itemize}

In summary, in our model the nova nucleosynthesis of $^7$Li is fully
described by three parameters: the delay time between the end of life
of the primary star and the first nova outburst ($\tau_{nova}$), the
number of binary systems which develop a nova ($N_{lithium}$) and the
total \livii\ produced by a nova in its lifetime $^{Li}Y_{Nova}$.

By comparison with the observations we explore the space of these three
parameters.  For the parameter $\tau_{nova}$, we investigate
timescales of 0, 1, 2 and 5 Gyr.  The nova bursts probably will
explode with different timescales depending on the mass loss of the
secondary star, but a quantitative lower limit, i.e. the shortest
timescale allowed, can be derived from the upper envelope of the data
in the [Li/H] vs [Fe/H] or [Li/H] versus time.

The other two parameters, the nova yields $^{Li}Y_{Nova}$ and the nova
rate $N_{lithium}$, are degenerate in terms of \livii\ production in
the model since the total \livii\ produced by a stellar generation is
proportional to their product. Thus, in our modelling we keep 
the rate $N_{lithium}$ fixed and vary only $^{Li}Y_{Nova}$. The comparison
with the data is an effective constraint to their combined
production. An estimation of uncertainties is obtained by applying a
change of production of 33\% to the standard value of $^{Li}Y_{Nova}$=
$1.8 \times 10^{-5}$ \msun.  We will also present a model with 
  more extreme enhancement $^{Li}Y_{Nova}$= $4.14 \times 10^{-5}$
  \msun~ that is able to recover the lithium abundance measured in
  meteorites.

  The parameter $N_{lithium}$ varies the number of nova systems
    in the model results and also the number of bursts expected
    nowadays, assuming that each nova system produces 10$^4$
    bursts. We take a value of $N_{lithium}$=0.03 that
  provides a rate of nova bursts compatible with the $20\,-\,30$ yr
  $^{-1}$ observed in the Galaxy at the present time
  \citep{Shafter97}.

Lithium is normally destroyed inside the stars at temperatures of few
million degrees and therefore the surface abundance is decreased by
the stellar mixing.  This process, i.e. the astration, is taken into
account in our model by assuming that \livii\ is destroyed completely
in stars of all masses.

A known \livii\ contribution comes from spallation of the cosmic rays
with the nuclei in the interstellar medium.  This production has a
rather small impact that is not sufficient to account for all the
observed lithium in our Galaxy.  In our model, we consider this
contribution starting from the observations of the stable isotope
$^9$Be \citep{ Molaro97b, Smiljanic09}. $^9$Be is produced only by
cosmic rays, and therefore by scaling this production it is possible
to evaluate also the lithium production by cosmic rays. We obtain this
relation for \livii :
$\log (Li/H) = -9.50 + 1.24 [Fe/H]$, starting from the relation
derived for $^9$Be by \citet{Smiljanic09}, and assuming as
scaling ratio \livii\ / $^9$Be $\sim$ 7.6 \citep{Molaro97b}.
Their results are obtained by
means of a linear fit applied to 73 stars belonging mostly to the halo
(39) and the thick disk (27); only 6 stars belong to the thin disk and
a star has equal chances to belong to the halo or to the thick disc.
Possibly, it would be more accurate for our task to have more data for
the thin disk.  However, to our knowledge these are the best results
available and they will have a limited impact - as we will see - on
the final results for the thin disk. In \citet{Smiljanic09}, a linear
fit is also computed for the thick disc. This could be specifically
used for our thick disk model, but given the almost negligible
difference, we keep the same prescription in both models.

Finally, also the contribution by AGB is considered. The theoretical
yields computed by \citet{Ventura13} are included in the chemical
evolution model for stars in the mass range 1\msun $<$M$<$6\msun.

In the context of this paper, we made the assumption that the initial
gas composition has \livii\ abundance of A(Li)=2.25. This assumption
is not critical as long as the reason which produces the disagreement
with the Planck \livii\ value does not depend on metallicity and
affects all the stellar populations in the same way.  To explore a
  possible solution for this disagreement, we also run a model with an
  initial A(Li)=2.6, so compatible with abundance of \livii\ derived
  by measurements of the CMB of Planck.

\subsection{Results for the thin disk}

\begin{figure*}[ht]
\includegraphics[width=120mm]{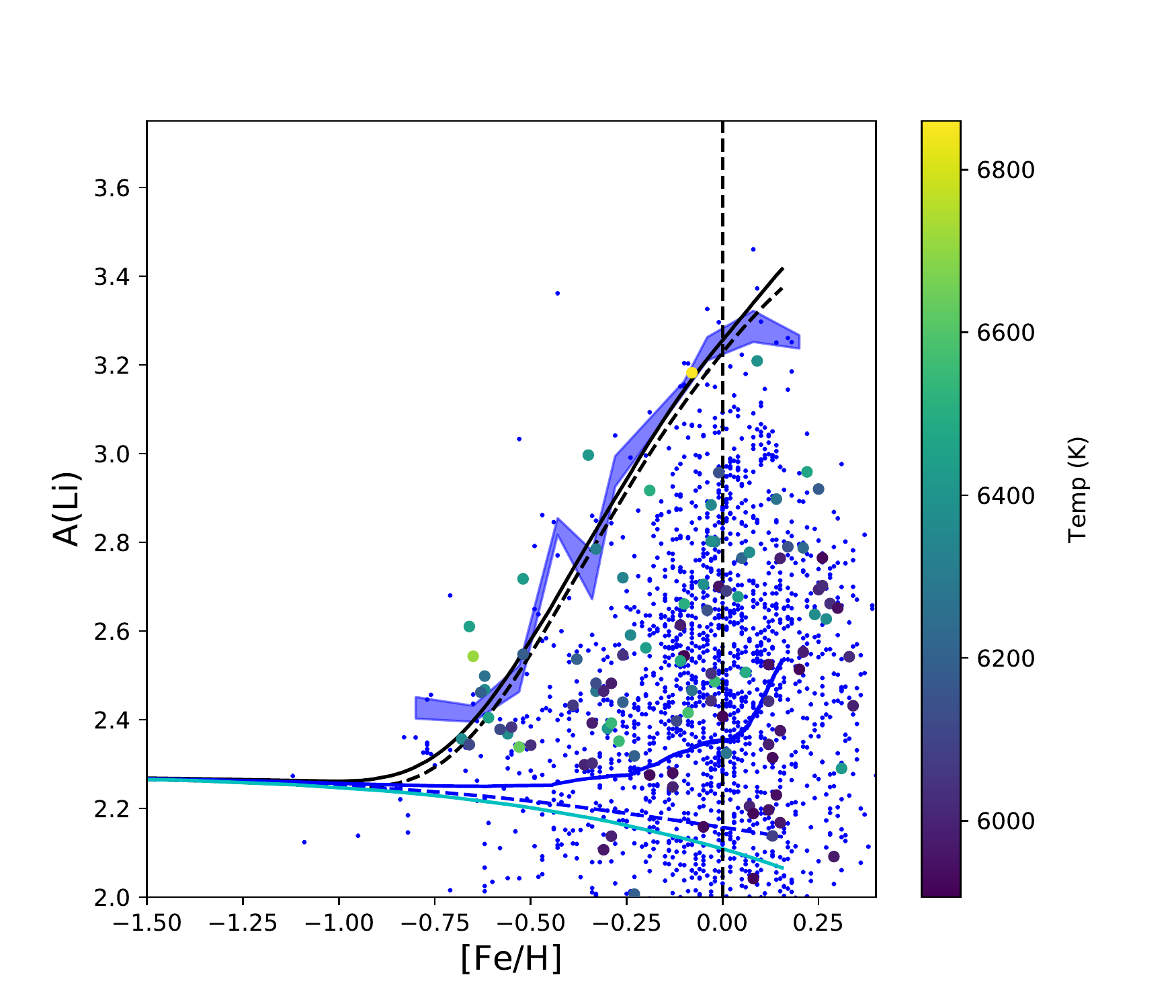}
\caption{\livii\ abundances vs [Fe/H] for the Galactic thin disk.  The
  blue dots are the \livii\ abundances of thin disk stars from the
  Ambre project. The larger symbols are the measurements of \livii\
  abundances for thin disk stars from \citet{Bensby18} color coded
  according to their stellar temperature. The shaded area highlights
  the upper envelope of the Ambre project data, defined in
  Sect.\ref{data_thin}. The impact of the different \livii\ factories
  are shown. Solid cyan shows only astration with no \livii\
  production; dashed blue is the production by AGB stars, solid blue
  the production by spallation; dashed black line the production by
  novae. The sum of all the three factories is shown as black solid
  line.  }
 \label{fig0}
\end{figure*}

\begin{figure*}[ht]
\includegraphics[width=120mm]{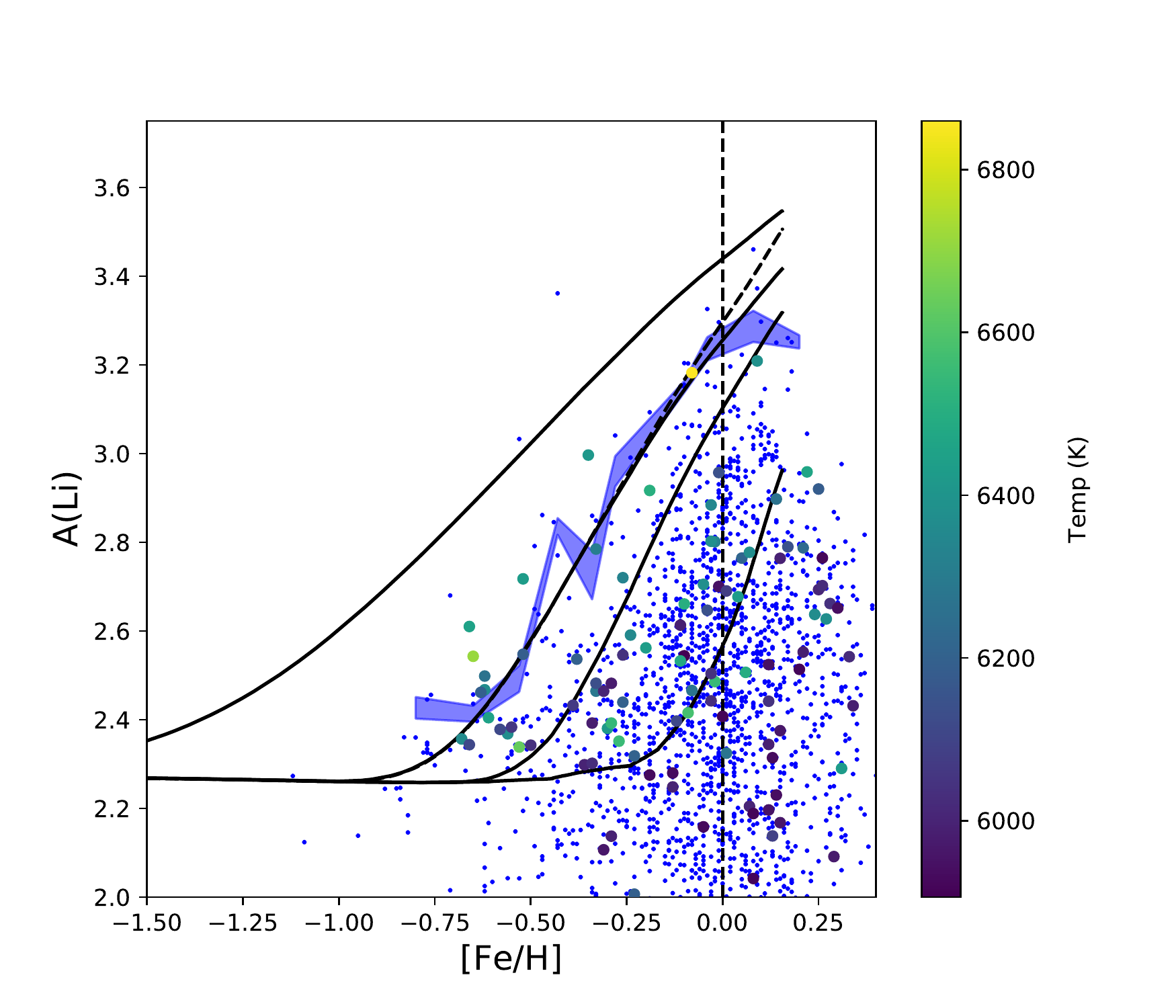}
\caption{\livii\ abundances vs [Fe/H] for the Galactic thin disk.  The
  observational data are the same of \ref{fig0}. Models with five
  different characteristic times ($\tau_{nova}$) of \livii\ enrichment
  are shown, from left to right: $\tau_{nova}$ = 0, 1, 2 and 5 Gyr.
  The dashed line shows the effect of changing the lower limit to
  $M_{ow}$=1.6 \msun, from $M_{ow}$=3 \msun, applied only to the
  model with $\tau_{nova}$=1 Gyr for simplicity.}
\label{fig1}
\end{figure*}

\begin{figure*}
\includegraphics[width=120mm]{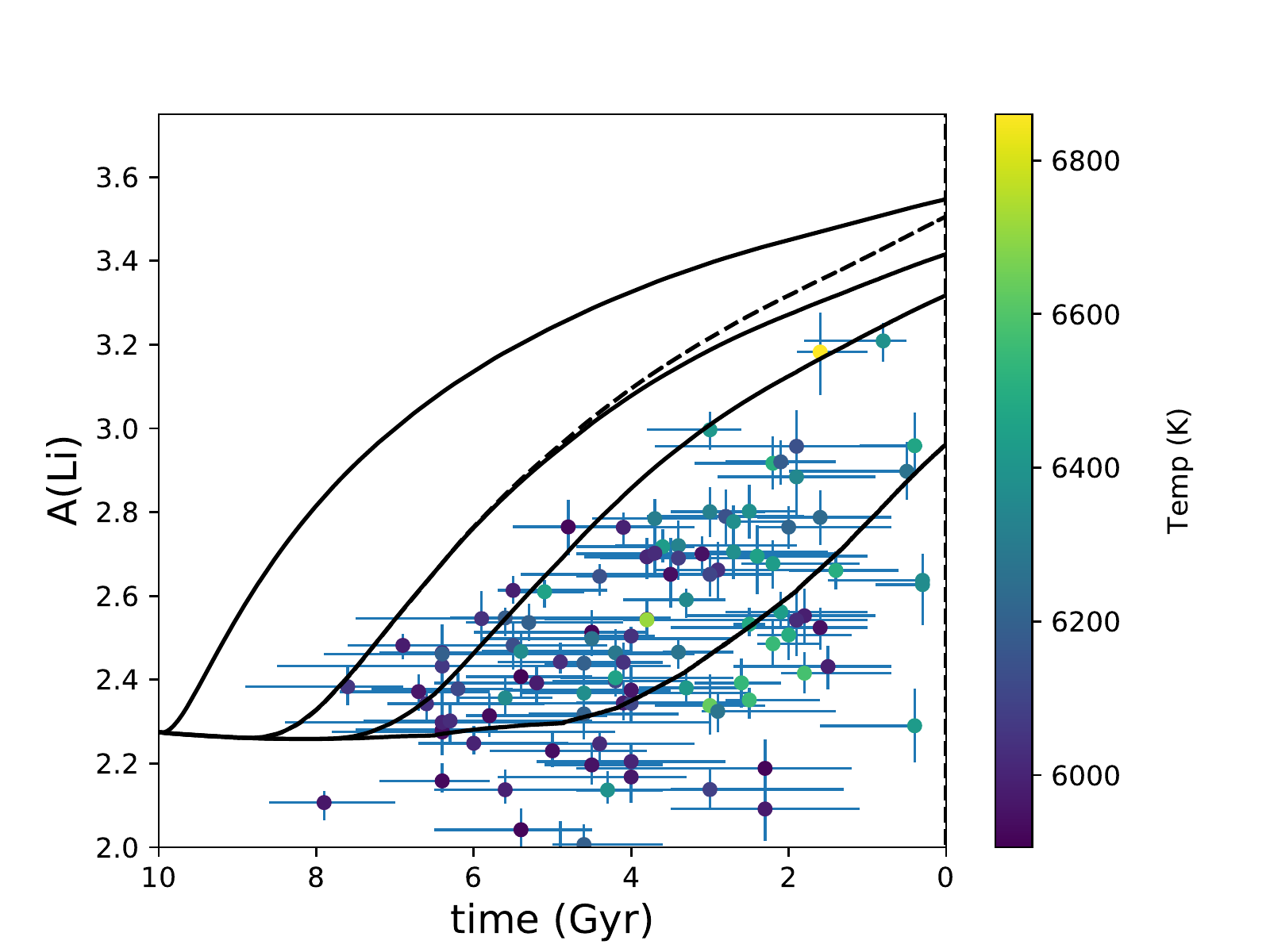}
\caption{\livii\ abundances vs time for the Galactic thin disk.  The
  circles show  the \livii\ abundances from \citet{Bensby18}
  with ages from \citet{Bensby14}.  Models with five different
  characteristic times of \livii\ enrichment are shown, from left to
  right: $\tau_{nova}$ = 0, 1, 2 and 5 Gyr. The dashed line presents
  the results of assuming a lower limit for the binary system
  $M_{ow}$=1.6 \msun~ as in Fig. \ref{fig1}. }\label{fig2}
\end{figure*}

\begin{figure*}
\includegraphics[width=120mm]{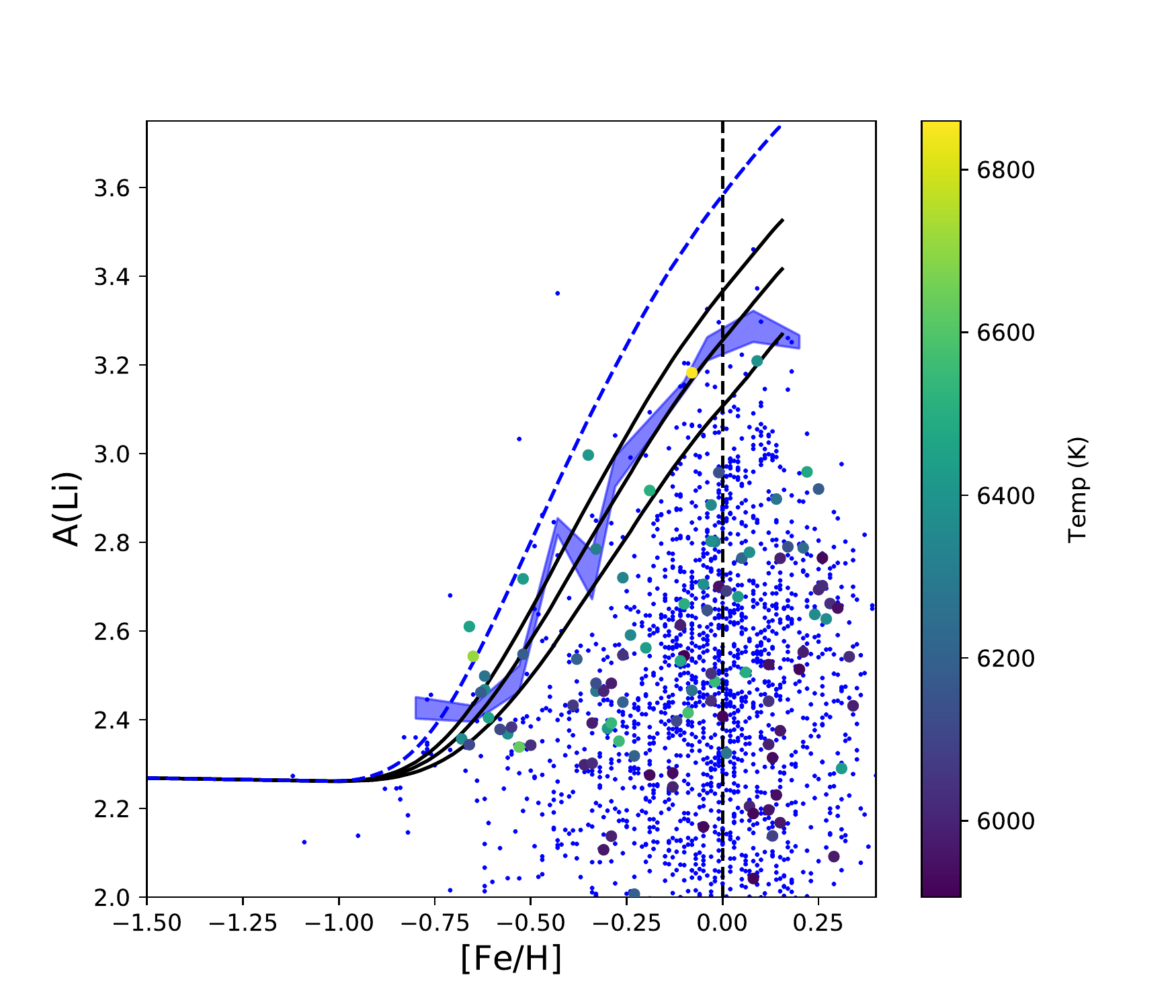}
\caption{ Same of Fig. \ref{fig1}. Effects of changing the parameter $^{Li}Y_{Nova}$. The 
  central black line is the model with the standard value of
  $^{Li}Y_{Nova}$, while the upper (lower) black line is
  obtained by increasing (decreasing) of 33\% the 
  $^{Li}Y_{Nova}$. The dashed blue line shows the model obtained increasing
  by a factor 2.3 the parameter $^{Li}Y_{Nova}$ .}\label{fig3}
\end{figure*}

\begin{figure*}
\includegraphics[width=120mm]{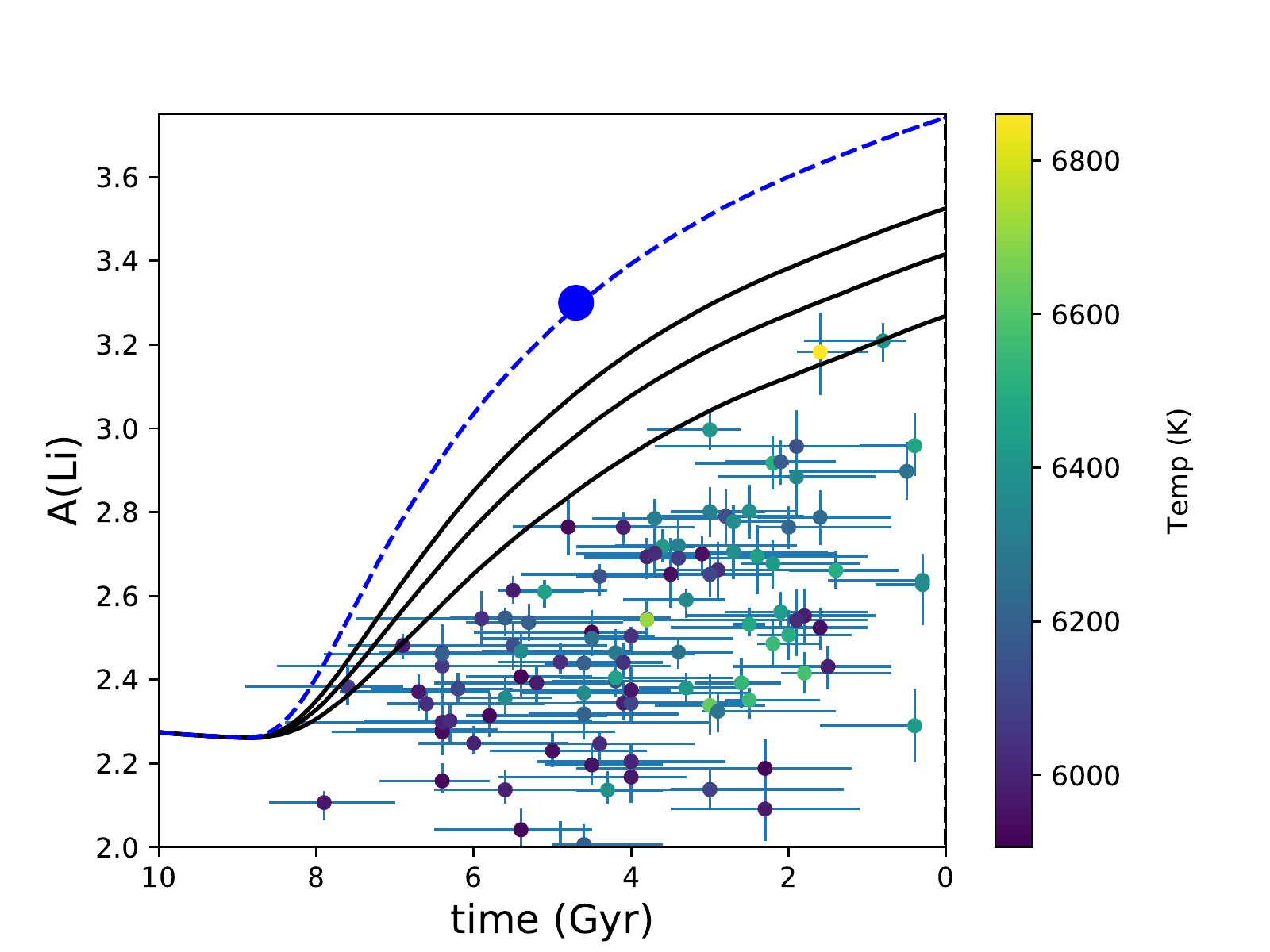}
\caption{ Same of Fig. \ref{fig3}. Effects of changing the $^{Li}Y_{Nova}$. The 
  central black line is the model with the standard value of
  $^{Li}Y_{Nova}$, while the upper (lower) black line is
  obtained by increasing (decreasing) of 33\% the 
  $^{Li}Y_{Nova}$. The dashed blue line shows the model obtained increasing
  by a factor 2.3 the parameter $^{Li}Y_{Nova}$ .}\label{fig4}
\end{figure*}

In this section we analyse the results obtained with our 
chemical evolution model for  the thin disk. 

In Fig. \ref{fig0} we show the thin disk chemical evolution with
different \livii\ factories: AGB stars, spallation of cosmic rays and
novae taken separately or all together. For illustrative purposes, in
the figure is also shown the extreme case where only astration is
taken into account, namely when there is no \livii\ production.

As displayed in Fig. \ref{fig0}, the AGB contribution to \livii\
increases the \livii\ by $\sim$0.05 dex.  Therefore, we confirm
  that the enrichment due to AGB stars does not influence the overall
  chemical evolution of \livii.  This was also a result of \citet{rom99,
  Romano01}.  We note that in the stellar evolution model
  considered \citep{Ventura13}, lithium production can be activated
  efficiently only in intermediate mass stars ($\ge $ 3\msun), and 
the possible contribution from extra mixing processes in
  low mass red giant branch or AGB stars is not included.

The spallation of cosmic rays starts to play a role from a
metallicity of [Fe/H]$\sim -$0.5, increasing by a factor of 2 (0.3
dex) the primordial level assumed here, and a factor of 3 (0.5 dex)
compared to the lithium destroyed by astration.  Thus, the total
amount of \livii\ is still dominated by a contribution from a
different source.  This was  the result by \citet{Prantzos12}.

Likely, the dominant source for \livii\ is the nova contribution. In
fact, the nova contribution assumed here with $^{Li}Y_{Nova}$ =
$1.8 \times 10^{-5}$ \msun~ and $\tau_{nova}$= 1 Gyr can reproduce
the observed values and the general behaviour of the
Galactic growth very well.

In Fig. \ref{fig1} are shown the \livii\ evolution results
considering four different delay times for the start of \livii\
enrichment, i.e. $\tau_{nova}$: 0, 1, 2 and 5 Gyr keeping fixed
$^{Li}Y_{Nova}$ = $1.8 \times 10^{-5}$ \msun~ per nova.  For
negligible delay-time (0 Gyr), the \livii\ abundances start to rise
almost immediately due to a very short time scale for the \livii\
enrichment.  In this case, the earliest effects are as short as 30Myr,
the lifetime of a 8\msun~ star. There are some sparse data points
close to this line and also above, nevertheless they are a negligible
fraction and probably do not reflect the general behaviour.  The
concordance with the bulk of  data requires a longer characteristic
timescale.  A timescale of 1 Gyr provides the best
match to the upper envelope, whereas 2 Gyr for $\tau_{nova}$ is
slightly too long. We could actually search for the precise timescale
to match the upper envelope, but given the uncertainties considered,
we prefer to keep 1 Gyr as best value for $\tau_{nova}$.  It is
interesting to note that a delay of 5 Gyr is largely missing the upper
envelope of the data, but still intercepts the stars with slightly
lower \livii\ . We can actually consider a more complex situation for
the lithum chemical evolution in which stars with a lower initial
value show their real initial lithium rather than a depleted value. In
this case longer time-scales for $\tau_{nova}$ could account for the
observed spread below the envelope.

We also studied the impact of changing the lower limit for the mass of
the binary system from 3 \msun~ to 1.6 \msun and this model is shown
in Fig. \ref{fig1} with a dashed line. The effects of a change
at the lower end of the masses appear evident on long timescales, and
indeed the trend of lithium above solar metallicity is steeper for the
model with $M_{low}=$1.6\msun. Since the data appear not to have such
steep trend, we use $M_{low}=$3\msun. We note that if we further
increase $M_{low}$, this will also impact the trend at intermediate
metallicity ($-$1$<$[Fe/H]$<$0) and will worsen the fit to the data.
A possible solution would be to have mass dependent yields, but - as
mentioned before - we prefer to keep modelling as simple as
possible.

In Fig. \ref{fig2}, we use the ages derived for stars in
\citet{Bensby14}, instead of the usual proxy for the ages: [Fe/H]. In
this case we cannot compute an upper envelope because of the low
number of stars, but the timescale $\tau_{nova}$ of 1 Gyr is a
suitable value also in the space A(Li) vs Age. The model
$\tau_{nova}=$ 2 Gyr appears the best in this plot, but we suspect
that this can be due to the low number of stars available with
ages. Still, this is the first time that chemical evolution model
results for lithium are shown against the ages of the stars and the
results are extremely good.

 In Fig. \ref{fig3}, we explore the change in the evolution
  obtained by changing the \livii\ yields.  A new curve (blue line) is
  obtained by multiplying the $^{Li}Y_{Nova}$ standard value by a
  factor 2.3.  We intend to study the impact of the variation of this
  parameter and we selected this extreme value to reproduce the
  solar Li abundance of 3.3 measured in meteorites formed at the time
  of solar system formation, 4.56 Gyr ago (cfr. Fig. \ref{fig4}). We
  note that this model overestimates the lithium abundance in
  all the thin disk stars, apart few extreme cases of AMBRE sample.
  Other two curves are instead obtained by a milder variation of 33\%
  to the standard value of $^{Li}Y_{Nova}$. The motivation of these
  new curves is to evaluate the variation needed to encompass the
  upper envelope of the AMBRE set of data, as shown in Fig. \ref{fig3}.
In Fig. \ref{fig3} we also note that the mild decrease of the yields
  mimics an increase in the $\tau_{nova}$ but the curves are not
  identical and they can be discriminated in the future by an
  increased number of good quality observations of stars. 

  In Fig. \ref{fig4} we show the same model results, but comparing
  A(Li) to ages.  The model with the mildly decreased yields
  can intercept all the stars in the \citet{Bensby18} sample with very high
  \livii\ abundance at given ages. This could be also obtained with a
  slightly longer $\tau_{nova}$, basically confirming the results in Fig.
  \ref{fig3} using ages instead of [Fe/H].

  In Fig. \ref{fig4} the model with the $^{Li}Y_{Nova}$ increased by a
  factor of 2.3 predicts a lithium abundance too high when compared to
  the entire \citet{Bensby18} sample and matches only the solar system
  abundance measured in meteorites (by construction). We present a
  possible explanation to this problem, the solar lithium problem, in
  Sect. \ref{Solarproblem}.

In summary, the model for the thin disk provides a $\tau_{nova}$ of 1
$\pm 0.5$ Gyr, and an effective nova yield $^{Li}Y_{Nova}$ of
$1.8 \pm 0.6 \times 10^{-5}$ \msun~ showing that, apart from a
relatively small contribution from cosmic rays and an even smaller
contribution from AGB stars, the bulk of Galactic \livii\ can be
consistently made by novae which are able to provide not only the
amount of \livii\ at the end of the evolution but also the rate of
increase of this element in the thin disk during its
chemical evolution.

\subsubsection{The solar lithium problem}\label{Solarproblem}

\begin{figure*}
\includegraphics[width=120mm]{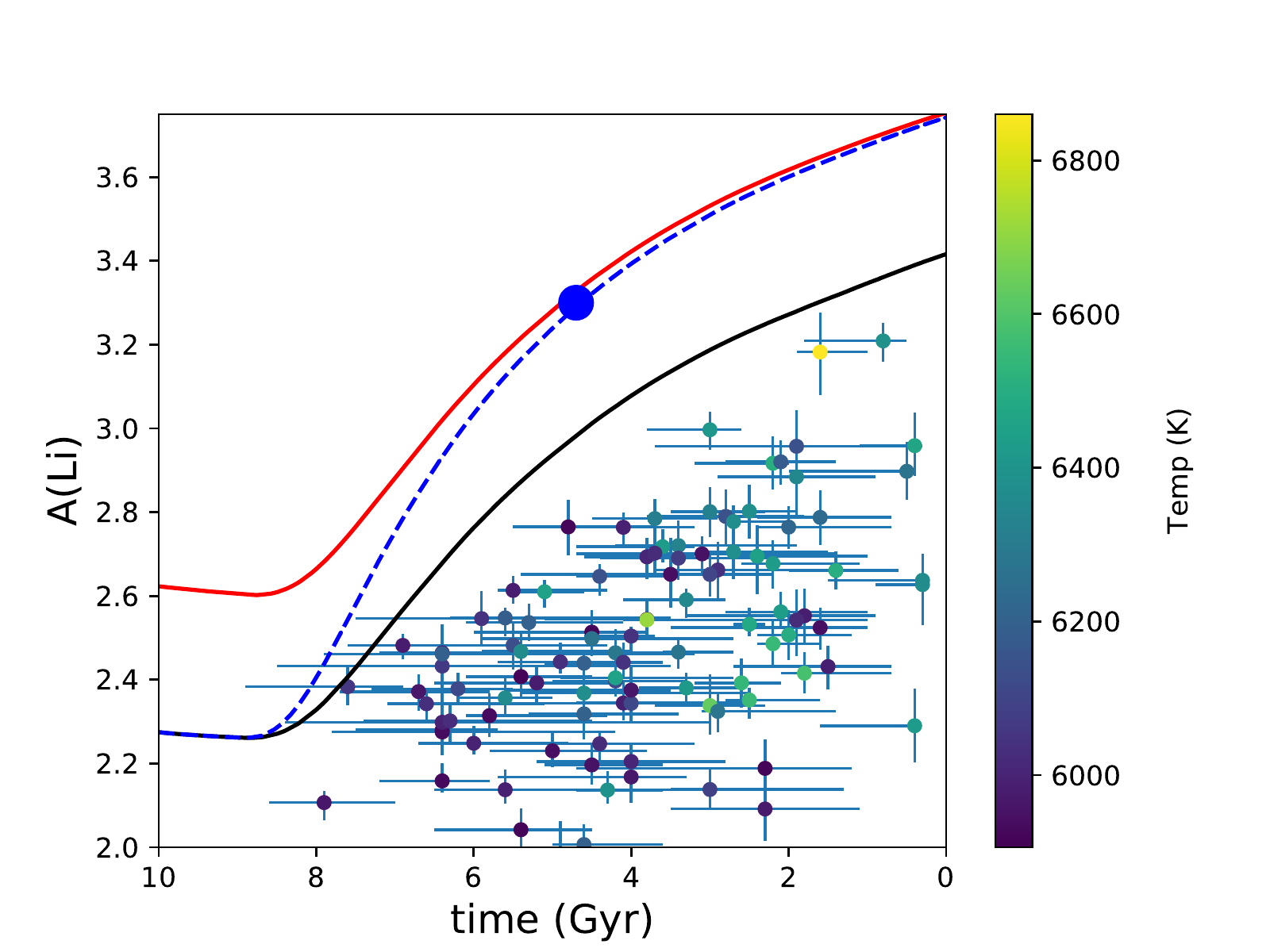}
\caption{ Same of Fig. \ref{fig3}. Effects of assuming a depletion
  during the pre main sequence phase. The solid line is the model with
  the standard value of $^{Li}Y_{Nova}$. The dashed blue line shows
  the model obtained assuming a $^{Li}Y_{Nova}$= $4.14 \times 10^{-5}$
  \msun. The red line is the model again with an increased
  $^{Li}Y_{Nova}$= $4.14 \times 10^{-5}$ \msun, but here we also assume
  an initial A(Li)=2.6.}\label{figdep}
\end{figure*}

As mentioned in the previous section, our model tuned for stellar
observations fails to match the solar value. Only by increasing the
effective yields by a factor of 2.3 the model reaches the solar
abundance measured in meteorites. However, the meteorites value and
the stellar determination may differ since they are obtained with
different techniques and at two different moments of the stellar life
and may not be directly comparable. In particular, the meteoritic
value is not affected by pre main sequence depletion, contrary to
stellar measurements \citep{Molaro12,Fu15,Thevenin17}.  In
Fig. \ref{figdep}, we present a possible solution. The model pre-main
sequence depletion (red line) is obtained increasing again by 2.3
times the yields from novae, so $^{Li}Y_{Nova}$=4.14 $\times 10^{-5}$
\msun, but starting from a primordial A(Li)=2.6 taken by the Big Bang
nucleosynthesis.  Interestingly, the meteorites value is matched and
the model is also able to reach the lithium abundance in TTauri stars,
extremely young objects.

Therefore, if we consider a simplified assumption that all stars
suffer a depletion of approximately 0.35 dex during the pre-main
sequence, the model falls back to the standard case (black line).  If
this is the case, the real effective yields from novae should be
approximately 2 times larger than the $^{Li}Y_{Nova}$ obtained from
stellar constraints and - according to our computation -$^{Li}Y_{Nova}$=
$4.14 \times 10^{-5}$ \msun.  We note that this value is still
within the uncertainties of the lithium production from novae
bursts.

\section{\livii\ Galactic Chemical evolution of the thick disk}

\subsection{ The Galactic thick disk}

The Milky Way, like almost two thirds of disk galaxies, has two major
disk components, the thin and the thick one with different histories
and formation \citep{Gilmore83}.  At the solar vicinity, the thick
disk dominates the stellar population at distances between 1 and 5 kpc
above the Galactic plane and it is generally accepted to be the oldest
part of the disk, i.e. older than $\approx$ 8-9 Gyr. 
The mechanism of formation of the thick disk is still debated.  The
proposed scenario includes vertical heating from infalling
satellites \citep{Quinn93,Villalobos08}, turbulent gas-rich
disk phase at high redshift \citep{Bournaud09, Forbes12}, 
and massive gas rich satellites \citep{Brook05}.
Formation of thick disks by radial migration was also proposed as a
mechanism by \citet{Schoenrich09}. This idea was challenged by
\citet{Minchev12}, demonstrating that migrators in N-body
models do not have any significant effect on disk thickening. Several
groups have now supported these findings in more recent
works \citep{Martig14,Grand16}, establishing this as a generic result
of disk dynamics.  In the most recently proposed model for the
formation of thick disks by \citet{Minchev15,Minchev17}, it is shown
that in galactic disks formed inside-out, mono-age populations (groups
of coeval stars) are well fitted by single exponentials and always
flare (the disk thickness increases with radius). In contrast, when
the total stellar density is considered, a sum of two exponentials is
required for a good fit, resulting in thin and thick disks which do
not flare.  Such a scenario explains why chemically- or age-defined
thick disks are centrally concentrated \citep{Bensby11,Bovy12}, but
geometrically thick populations in both observations of external
edge-on galaxies \citep{Comeron12} and in the Milky Way
\citep{Robin96, Juric08} extend beyond the thinner component.  From
this scenario, thick disk stars at the solar vicinity are the oldest
mono-age populations.


\subsection{\livii\ in   thick disk stars}

The first attempt to measure Li in thick disk stars was made by
\citet{Molaro97}, who measured Li in seven metal poor thick disk stars
and found an abundance of Li consistent with that of the halo stars.
\citet{rom99} used the kinematical properties to separate the thin
disk from the thick disk stars. In their sample there is also a thick
disk star BD+01 3421 with [Fe/H]=$-$0.5, but with a lithium abundance
of A(Li)=2.11 which is much lower than the Li abundance of thin disk
stars of similar metallicity.  The lack of increase increase of Li
abundance among the thick disk stars was established by
\citet{Ramirez12} on a larger data sample showing that the maximum
thick disk lithium abundances remain close to the Spite plateau
regardless of their metallicity.  \citet{Delgado15} and
\citet{Bensby18} even suggested that the thick disk lithium abundances
decrease with increasing metallicity.  \citet{Guiglion16} used the
AMBRE catalogue composed of $363$ stars and found that the highest
lithium abundances seem to increase slightly with [Fe/H], from about
Li = 2.0 up to 2.4 dex at [Fe/H]=-0.3.  In the AMBRE sample the
distribution of Li abundances in the thick disk clusters is at
$\sim2.2$ and $\sim1.2\,$dex, respectively. The higher value seems to
correspond to an extension of the Spite plateau, while the lower
values are likely due to lithium destruction in stellar interior.
\citet{Fu18} investigated the Li enrichment in the sample of 1399
stars of the CES iDR4 sample.  By means of the chemical division
proposed by \citet{Adibekyan12}, they identified 73 stars belonging to
the thick disc. However, these generally have a relatively higher
abundance of A(Li) $\approx$ 2.4, and possibly a modest rise at the
highest metallicity.

\subsection{\livii\ evolution model  for the thick disk}

In this Section, we probe the hypothesis that novae are the main Li
source also in the thick disk by using the parameters that reproduce
at best the lithium evolution in thin disk.  We use the standard
  model and we do not test the pre-main sequence depletion model. In
  fact, given the available constraints, we prefer to adopt the model
  able to reproduce the stellar observations.  In our model the thick
and thin disks evolve independently from each other without any
exchange of gas or stars.  Our approach is similar to that followed by
\citet{Grisoni17} for the [$\alpha$/elements].  Instantaneous
recycling is relaxed and the stellar lifetimes are taken into account
as for our thin disk model.

The thick disk is characterised by a more intense star formation
history than the thin disk. There  is a quicker  evolution with a stronger
efficiency ($\nu$=3) and a shorter time scale of 0.1 Gyr for infall.
The chemical evolution  lasts for 2 Gyr, but with only residual
star formation after approximately 1.2 Gyr. These parameters are well
suited to account for  the oldest mono-age populations, following the
scenario presented in \citet{Minchev17}.
For the rest, the model is identical to that of the thin disk.

\subsection{Results for the thick disc}
\begin{figure*}
\includegraphics[width=180mm]{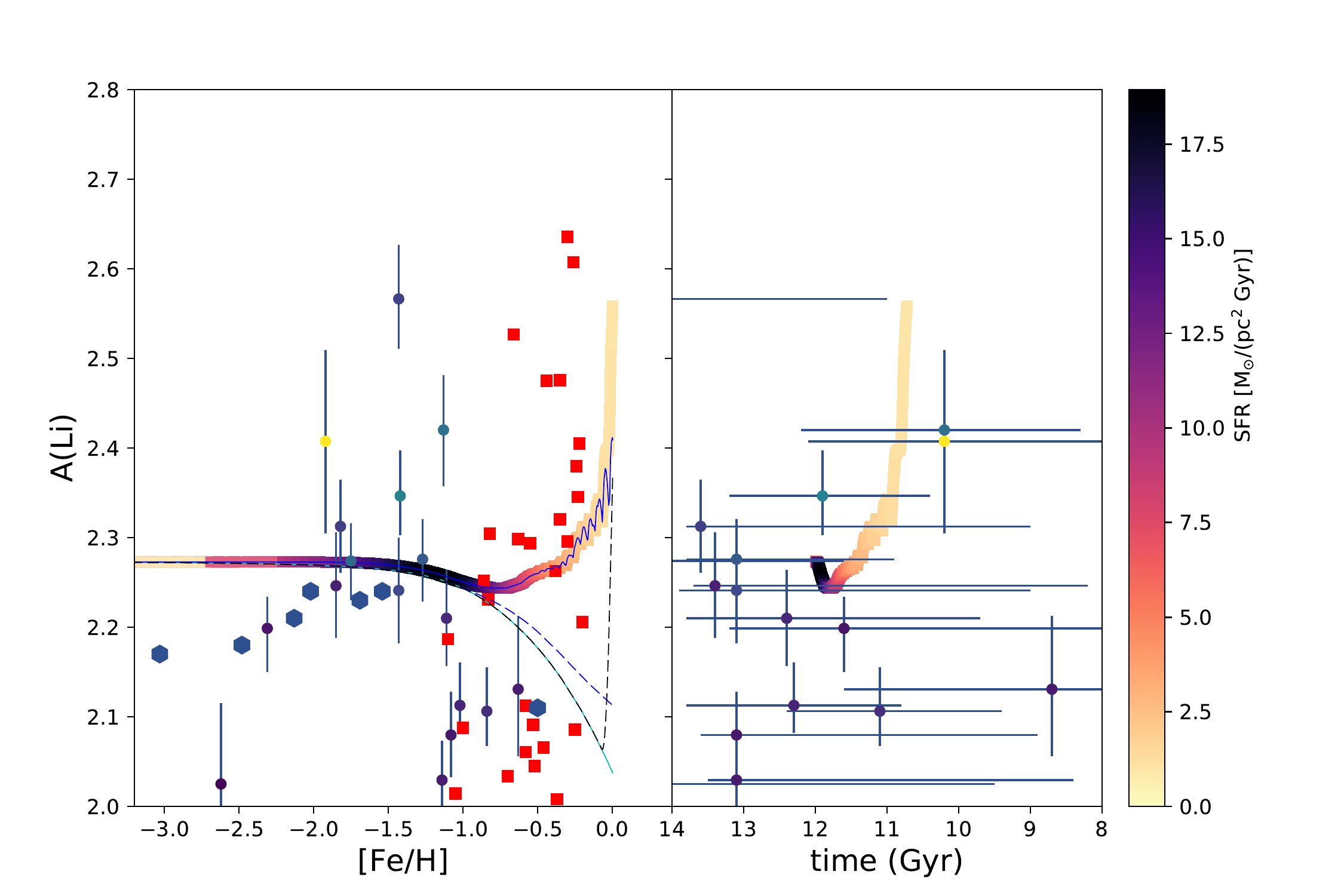}
\caption{Results for the Galactic thick disc. Left panel: \livii\
  abundances vs [Fe/H].  The red squares are the stars from Ambre
  project; the circles are thick disk stars from \citet{Bensby18}
  color coded according to their stellar temperature (see colorbar in
  previous figures); blue hexagons are from \citet{Molaro97}. The
  thick line shows the model results and it is color coded depending
  on the star formation rate. The thinner lines show the model results
  for: no \livii\ production (cyan), only novae (black dashed), only
  AGB (blue dashed) and only spallation (blue). Right panel: \livii\
  abundances vs time. The data are from \citet{Bensby18} while the
  thick line shows the model results and it is color coded depending
  on the star formation rate as in the left panel.}\label{fig5}
\end{figure*}
 
Our model with $\tau_{nova}$ of 1 Gyr, and with the same Li yields
derived for the thin disk stars predicts no increase in the \livii\
abundance for thick disk stars up to [Fe/H]=$-$0.2. This is because
the time of the evolution of the thick disk is shorter than the
characteristic time in which novae start to contribute with Li.  At
variance with the thin disk evolution our model for the thick disk
predicts no Li increase in excellent agreement with the observational
result.  Actually, if we consider only novae production, \livii\ should
decrease and is kept about constant under the action of spallation
processes that is the only active nucleosynthetic process in the
thick disk.  \citet{Bensby18} for the first time provide an age
estimation for the thick disk stars.  These ages are compatible with
an old thick disk as we have assumed in our model. We note however
that some stars have relatively younger ages (and fewer ones
older). We cannot explain the ages of this younger population.  They
can be due to a false thick disk identification or a problem in
measuring the ages.  Our model produces a thick disk in a relatively
short time-scale of 2 Gyr. Thus, assuming that the formation of the
thick disk started 12 Gyr ago, its star formation was over about 10 Gyr ago.
Following \citet{Bensby18} we consider {\it bona fide} thick disk
stars the ones with an age greater than 8 Gyr. The result of our
model for the thick disk in the A(Li) vs age space is shown in
Fig. \ref{fig5}.  Given the uncertainties in the age determination
there is a substantial agreement with the model.

At the high metallicity end [Fe/H]=$-$0.2 our model predicts a sharp
increase of the \livii\ abundance reaching values up to
A(Li)$\sim$2.6. This is because novae start to contribute
significantly. However, the very mild SFR that characterises this phase
would effectively mark the end of stellar formation of the thick disk.
Our threshold for stellar formation is conservatively set at 1 \msun
$pc^{-2}$, while, for comparison, was set to 4 \msun $pc^{-2}$ in
\citet{Chiappini97} for the thick-halo phase.  It is not clear whether
there is a real chance to find stars of the thick disk in this area,
but if they are, according to our model, they should show relatively
higher Li abundances than the average of thick disk stars.

\section{Conclusion}

The idea that nova systems could produce lithium dates back to the mid
70s by \citet{arn75} and \citet{sta78}.  They applied to the novae the
idea put forward by Cameron-Fowler \citep{cam71} to explain high
\livii\ abundances in some luminous red giants.  This possibility was then
incorporated in the Galactic chemical \livii\ evolution models by
\citet{D'Antona91} and \citet{rom99} together with other possible Li
sources to explain the \livii\ behaviour. However, the non detection
of the \liviii\ line in the novae outbursts led these authors to
favour low mass red giants as the most likely Li source
\citep{Romano01,Prantzos17}.  Observational evidence has been found
only recently with the systematic detection of the mother nuclei
$^{7}$Be in the post-outburst spectra of classical novae by \citet{taj15},
\citet{taj16}, \citet{Molaro16}, \citet{ Izzo18} and
\citet{Selvelli18}.  The observed \bevii\ decays into \livii\ with a
mean-life of 53 days and all observed \bevii\ decays into \livii.
Moreover, the \bevii /H has been measured with abundances that are
higher than the novae theoretical models and provide overproduction factors
between 4 and 5 dex over the meteoritic abundance of \livii.  The
total amount of \livii\ produced by a nova is quite uncertain since it
depends on the ejected mass and on the number of outbursts a novae
experiences in its whole life.
 
In this paper we have revised the \livii\ evolution by considering
these most recent observations.  By means of a detailed chemical
evolution model for the thin disk we have shown that nova systems
start to enrich the interstellar \livii\ with a delay of about 1 Gyr.
In order to match the increase of Galactic \livii\ observed in the
thin disk, an effective yield of $1.8 (\pm 0.6) \times 10^{-5}$\msun~
of \livii\ per nova in a whole life is required.  It is quite
remarkable that this value is consistent with what inferred in novae
outbursts.  Considering that there is an average \livii\ mass ejection
of $10^{-9}$\msun~ in a nova event and that there are about 10$^4 $
events per nova, we obtain a yield of $\approx 10^{-5}$\msun~ per
nova, which is remarkably close to our model derived yields
considering the uncertainties in all parameters.  This model is
  tuned to reproduce the Li observations in stars and fails to
  reproduce the meteoritic-solar value 4.56 Gyrs ago. However, it has
  been suggested that stellar abundances could have been affected by
  pre-main sequence destruction \citep{Molaro12, Fu15, Thevenin17}. In
  this context, we are able to reproduce the stellar trend increasing
  by two the standard effective yields, starting from the primordial
  value inferred by big bang nucleosynthesis and assuming a depletion
  of $\sim$0.35 dex during the pre-main sequence. This pre-main sequence
  depletion model matches the lithium abundance measured in the
  meteorites and young TTauri stars.

We have also applied the parameters derived from the thin disk \livii\
evolution to a model for the thick disk.  In this way we have shown
that assuming the classical novae as the main \livii\ factory we can
also explain the absence of any \livii\ enhancement observed in the
thick disk stars. We have found that the \livii\ does not decrease due
to astration thanks to spallation processes, that is the only active
nucleosynthetic process in the thick disk.  Thus, the \livii\
abundance can be used as a test to confirm or to reject the belonging
to the thick disk population.  This can be considered as a new
criterium to be used in combination with the more popular
$\alpha$-knee.

Extrapolating these results to the dwarf galaxies satellites of our
Milky Way, we expect the \livii\ abundances in stars to show an
enhancement in \livii\ only for those galaxies with a star formation
history long enough to provide novae products to contribute to the gas
enrichments. Thus, no \livii\ abundance above A(Li)= 2.3 is expected
in very small objects as ultra faint galaxies which evolve in short
time scales relative to the novae evolution. An important test will be
the Galactic Bulge which is likely to fall into this category,  at
  least its older and metal-poor population \citep{Cescutti18}.  We
intend to extend our prediction also to this structure of the Milky
Way as soon as we have new \livii\ abundance determinations in stars
belonging to this component.

 \section*{Acknowledgments}

 GC and PM thank Thomas Bensby, Xiaoting Fu and Guillaume Guiglion for
 sharing their data and useful discussions. GC also thanks Ivan Minchev
 for his advice in the section of the thick disk. G.C. acknowledges
 financial support from the European Union Horizon 2020 research and
 innovation programme under the Marie Sk\l odowska-Curie grant
 agreement No.  664931.  This work has been partially supported by the
 the EU COST Action CA16117 (ChETEC).

\label{lastpage}

\bibliography{litio}

\end{document}